\documentclass[11pt]{article}

\pdfoutput=1

\usepackage[T1]{fontenc}
\usepackage[latin9]{inputenc}
\usepackage[a4paper]{geometry}
\usepackage[active]{srcltx}
\usepackage{amsmath}
\usepackage{amssymb}
\usepackage{esint}

\makeatletter

\usepackage{jheppub}

\usepackage{etoolbox}
    
    \patchcmd{\maketitle}{\@fpheader}{}{}{}
    
\usepackage{aecompl}

\usepackage{amsfonts}
\usepackage{bbm}

\newcommand{\eq}[2]{\begin{equation} #1 \label{#2} \end{equation}}
\DeclareMathOperator{\extdm}{d}
\newcommand{\extd}{\extdm \!}

\title{Log corrections to entropy of three dimensional black holes with soft hair} 
\author[a,b,c]{Daniel Grumiller,}
\author[b]{Alfredo Perez,}
\author[b]{David Tempo,}
\author[b,c]{and Ricardo Troncoso}
\affiliation[a]{Institute for Theoretical Physics,
 TU Wien, Wiedner Hauptstr.~8-10, 1040 Vienna, Austria}
\affiliation[b]{Centro de Estudios Cient\'{i}ficos (CECs), Av. Arturo Prat 514, Valdivia,
Chile}
\affiliation[c]{Galileo Galilei Institute for Theoretical Physics, Arcetri, Italy}

\emailAdd{grumil@hep.itp.tuwien.ac.at}
\emailAdd{aperez@cecs.cl}
\emailAdd{tempo@cecs.cl}
\emailAdd{troncoso@cecs.cl}
\subheader{\begin{flushright}\textrm{CECS-PHY-17/03}\\\textrm{TUW--17--03}\end{flushright}}

\abstract{We calculate log corrections to the entropy of three-dimensional black holes with ``soft hairy'' boundary conditions. Their thermodynamics possesses some special features that preclude a naive direct evaluation of these corrections, so we follow two different approaches. The first one exploits that the BTZ black hole belongs to the spectrum of Brown--Henneaux as well as soft hairy boundary conditions, so that the respective log corrections are related through a suitable change of the thermodynamic ensemble. In the second approach the analogue of modular invariance is considered for dual theories with anisotropic scaling of Lifshitz type with dynamical exponent $z$ at the boundary. On the gravity side such scalings arise for KdV-type boundary conditions, which provide a specific 1-parameter family of multi-trace deformations of the usual AdS$_3$/CFT$_2$ setup, with Brown--Henneaux corresponding to $z=1$ and soft hairy boundary conditions to the limiting case $z\to 0^+$. Both approaches agree in the case of BTZ black holes for any non-negative $z$. Finally, for soft hairy boundary conditions we show that not only the leading term, but also the log corrections to the entropy of black flowers endowed with affine $\hat{u}\left(1\right)$ soft hair charges exclusively depend on the zero modes and hence coincide with the ones for BTZ black holes.}

\makeatother

\begin{document}
\maketitle \flushbottom


\section{Introduction}\label{se:1}

The Bekenstein--Hawking (BH) entropy law 
\begin{equation}
S_{\textrm{\tiny BH}}=\frac{A}{4G}\label{eq:log1}
\end{equation}
is currently the closest template to experimental data available to
rule out (or confirm) putative theories of quantum gravity, in the
sense that any theory claiming to quantize Einstein gravity but failing
to reproduce the BH entropy \eqref{eq:log1} in the classical limit
could be discarded. Like the leading, classical, term \eqref{eq:log1}
also log corrections to the BH entropy law depend only on the validity
of the semi-classical approximation and generically take the form
(see \cite{Carlip:2000nv,Sen:2012dw} and references therein) 
\begin{equation}
S=S_{\textrm{\tiny BH}}-q\,\ln S_{\textrm{\tiny BH}}+\dots\label{eq:log2}
\end{equation}
where the ellipsis denotes terms of order unity that we ignore in the present work. The numerical coefficient $q$ depends on the dimension, the thermodynamical ensemble and the matter content.

Quantum completions of Einstein gravity can provide microscopic derivations
for $q$, which sometimes disagree \cite{Kaul:2000kf} with the macroscopic
predictions from semi-classical Einstein gravity \cite{Solodukhin:1994yz,Fursaev:1994te,Mann:1997hm,Page:2004xp,Sen:2012dw}
and sometimes agree with it \cite{Sen:2012dw,Pathak:2016vfc}. Assuming
that one trusts in quantum field theory and Einstein gravity, one
can use this procedure to discard certain quantum gravity proposals.
Thus, it is of interest to calculate $q$ for various black holes
in order to have a data base for semi-classical checks.

In the case of three-dimensional pure Einstein gravity the BTZ black holes \cite{Banados:1992wn,Banados:1992gq} receive no log corrections, $q=0$, in the grand canonical ensemble \cite{Carlip:1994gc,Carlip:2000nv,Sen:2012dw}; transforming to the microcanonical one yields $q=3$ for Brown--Henneaux boundary conditions (bc's) \cite{Brown:1986nw}.

This can be seen as follows. Macroscopic corrections from thermal
fluctuations to the microcanonical entropy to leading order acquire
log corrections from left and right movers with energies $E_{\pm}$,
so that (see e.g. \cite{Landau:1980mil})
\begin{equation}
S_{\text{mic}}=S_{+}+S_{-}-\frac{1}{2}\ln\left(C_{+}T_{+}^{2}\right)-\frac{1}{2}\ln\left(C_{-}T_{-}^{2}\right)\,,\label{eq:Landau}
\end{equation}
with left and right entropies $S_{\pm}=\pi\sqrt{\frac{\ell}{G}E_{\pm}}$, left and right temperatures $T_{\pm}=T\left(1\pm\ell\Omega\right)^{-1}$ and left and right specific heats $C_{\pm}=\partial E_{\pm}/\partial T_{\pm}$. Thus, since the relationship between $E_{\pm}$ with $T_{\pm}$ for the BTZ black hole with Brown--Henneaux bc's is known to be given by
\begin{equation}
E_{\pm}=\frac{\ell\pi^{2}}{4G}T_{\pm}^{2}\;,\label{eq:StefanBoltzmannlaw-1}
\end{equation}
the microcanonical entropy in \eqref{eq:Landau} reduces to
\begin{equation}
S_{\textrm{\tiny mic}}^{\left(1\right)}=S_{+}+S_{-}-\tfrac{3}{2}\ln\big(S_{+}S_{-}\big)+\dots\label{microentropy-1}
\end{equation}
Therefore, following \cite{Sen:2012dw}, the log correction ($q=3$) is twice as large as the contribution from each chiral copy ($q=\tfrac32$). This result follows from the scaling $S_{\pm}\sim S_{\textrm{\tiny BH}}$ with an order unity proportionality constant that is state-dependent, but irrelevant in the semi-classical approximation, as long as we are not close to extremality. In the near extremal case one of the $S_{\pm}$ would be much smaller than $S_{\textrm{\tiny BH}}$ so that the proportionality constant in the scaling relation $S_{\pm}\sim S_{\textrm{\tiny BH}}$ cannot be neglected. We are going to assume that we are sufficiently away from extremality in the rest of the paper, apart from the concluding section.

Alternatively, the log corrections for the entropy of the BTZ black hole with Brown--Henneaux bc's can also be obtained from a direct microscopic computation using the grand canonical ensemble, which exploits the modular invariance of the dual theory. As shown in \cite{Carlip:1994gc,Sen:2012dw}, the log corrections obtained in this way agree with the result $q=3$. 

Hereafter, we focus on obtaining the log corrections to the black hole entropy for a different class of~ ``soft hairy'' bc's \cite{Afshar:2016wfy, Afshar:2016kjj} (also known as Heisenberg bc's \cite{Grumiller:2016pqb}). In addition to BTZ black holes their spectrum also includes ``black flowers'', describing stationary solutions whose event horizon is not spherically symmetric. The ripples at the horizon were shown to carry $\hat{u}\left(1\right)$ affine global charges that correspond to soft hair in the sense of \cite{Hawking:2016sgy, Hawking:2016msc} since they commute with the Hamiltonian. A remarkable feature of this new set of bc's is that the entire spectrum is compatible with regularity of the Euclidean solutions, independently of the precise value of the global charges including the mass and the angular momentum. In other words, the bc's in \cite{Afshar:2016wfy,Afshar:2016kjj} are such that the energy does not depend on the temperature at the equilibrium. Indeed, the inverse specific heat identically vanishes, which leaves no room for thermal energy fluctuations. Consequently, the log corrections to the black hole entropy cannot be obtained from a naive direct evaluation of \eqref{eq:Landau}.

In our present work we resolve this issue by calculating these corrections in two different ways, a macroscopic one that exploits the transformation of the density of states with a Jacobian under redefinition of energy, and a microscopic one that exploits a limiting case of a high/low temperature duality for theories with anisotropic scaling of Lifshitz type. Both lead to the same final result for the numerical coefficient $q_{\textrm{\tiny sh}}$ in \eqref{eq:log2} (the subscript stands for ``soft hairy''), namely
\eq{
q_{\textrm{\tiny sh}} = 1\,.
}{eq:mainresult}

This paper is organized as follows. In section \ref{sec:Log-corrections-from} we perform the computation of the log correction to the black hole entropy which exploits the fact that BTZ belongs to both spectra corresponding to Brown--Henneaux and soft hairy bc's. The result is then extended to a wider class of KdV-type bc's, which contains both Brown--Henneaux and the soft hairy ones as particular cases. 
In section \ref{sec:Microscopic-log-corrections} we carry out a microscopic computation following a different strategy. It makes use of an analogue of modular invariance for dual theories with anisotropic scaling of Lifshitz type, where the dynamical exponent can be used as a regulator in order to obtain the log corrections to the entropy of black holes with soft hairy bc's.  Section \ref{sec:Logarithmic-corrections-for-soft-hairy} is devoted to the case of black flowers endowed with affine $\hat{u}\left(1\right)$ soft hair charges. In section \ref{sec:outlook} we provide a brief outlook to possible generalizations.

\section{Log corrections from a suitable change of thermodynamic ensemble}\label{sec:Log-corrections-from}

\subsection{Soft hairy and Brown--Henneaux boundary conditions}\label{subsec:Soft-hairy-and}

As recently shown in \cite{Afshar:2016wfy,Afshar:2016kjj}, the BTZ
black hole also belongs to the spectrum of a new class of soft hairy
bc's. In particular, these bc's differ
from the ones of Brown and Henneaux by the choice of Lagrange multipliers
at infinity. Consequently, left and right energies for a given black
hole configuration, whose outer and inner horizons are determined
by the zero modes of the Virasoro generators ${\cal L}_{\pm}$, depend
on the precise choice of bc's.

In the case of Brown--Henneaux, left and right energies are given by
\begin{equation}
E_{\pm}^{\left(1\right)}=\frac{\ell}{16G}{\cal L}_{\pm}\,,\label{eq:BH-Energy}
\end{equation}
while for the soft hairy bc's they read
\begin{equation}
E_{\pm}^{\left(0\right)}=\frac{\ell}{8G}\sqrt{{\cal L}_{\pm}}\,.\label{eq:SH-Energy}
\end{equation}

In the context of black hole thermodynamics, inequivalent
choices of Lagrange multipliers at infinity translate into different
ways of fixing the chemical potentials at the boundary (see e.g.~\cite{Henneaux:2013dra}),
which means that we are dealing with the same object, but embedded
into different thermodynamic ensembles. Therefore, the log
corrections to the entropy of BTZ black holes that fulfill soft hairy
bc's can be obtained from the corresponding corrections
of the ones subject to Brown--Henneaux bc's given by
\eqref{microentropy-1}. This can be seen as follows. The microscopic
density of states in both ensembles are related by a Jacobian factor (see e.g. \cite{Pathak:2016vfc})
\begin{equation}
\rho\left[E_{\pm}^{\left(0\right)}\right]=\frac{\partial E_{\pm}^{\left(1\right)}}{\partial E_{\pm}^{\left(0\right)}} \, \rho\left[E_{\pm}^{\left(1\right)}\right]\,.\label{eq:jacoibin}
\end{equation}
Hence, the corrections we look for can be obtained from taking the logarithm of the microscopic density of states \eqref{eq:jacoibin}, which by virtue of Eq.~\eqref{microentropy-1} yields 
\eq{
S_{\pm}^{\left(0\right)}=S_{\pm}-\tfrac{3}{2}\ln S_{\pm} + \ln \frac{\partial E_{\pm}^{\left(1\right)}}{\partial E_{\pm}^{\left(0\right)}} \,.
}{eq:whyhasthisnolabel}
Taking into account the definitions of left and right energies in both ensembles, given by \eqref{eq:BH-Energy} and \eqref{eq:SH-Energy}, the leading term of the black hole entropy and its log corrections reduce to 
\begin{equation}
S_{\text{mic}}^{\left(0\right)}=S_{+}+S_{-}-\tfrac{1}{2}\ln\big(S_{+}S_{-}\big)\;.\label{eq:microentropy-0}
\end{equation}
Therefore, according to \eqref{eq:log2} and with the same caveat regarding non-extremality as below \eqref{microentropy-1}, one finds that the log
corrections to black hole entropy with soft hairy bc's
corresponds to $q_{\textrm{\tiny sh}}=1$, as announced in \eqref{eq:mainresult}.

\subsection{Boundary conditions of KdV-type}

Log corrections to black hole entropy can also be obtained along the lines above for a wider class of bc's. Here we focus on KdV-type bc's \cite{Perez:2016vqo} which contain both Brown--Henneaux and soft hairy bc's as particular cases. They are characterized by the fact that the lapse and shift functions in an ADM foliation are field dependent, being determined at the boundary by the variation of two independent functionals with respect to the dynamical fields. These functionals can be chosen as any of the possible Hamiltonians that span the KdV hierarchy, labeled by an integer $k$ called ``KdV label''. 

The asymptotic structure that arises turns out to be a very interesting one. The global gravitational excitations (boundary gravitons) fulfill the field equations of the $k$-th representative of the KdV hierarchy, so that they possess an anisotropic scaling of Lifshitz type with dynamical exponent $z=2k+1$. The isotropic case ($z=1$) reduces to the Brown--Henneaux bc's for which the boundary gravitons are described by chiral movers. It was also shown in \cite{Perez:2016vqo} that the KdV hierarchy can be extended in a non-trivial way if the corresponding Hamiltonians, $H_{\pm}^{\left(z\right)}$, are allowed to depend non-locally on the dynamical fields. Remarkably, the simplest example of this sort precisely corresponds to the soft hairy bc's of \cite{Afshar:2016wfy,Afshar:2016kjj} whose scaling properties correspond to $z=0$ (or equivalently, to a negative half-integer label $k=-1/2$). The spectrum of solutions then becomes radically different for distinct choices of $z$ that define the bc's. Nonetheless, it is simple to see that the BTZ black hole 
is always contained within all possible spectra.

The geometry of BTZ black holes with KdV-type of bc's,
specified by $H_{\pm}^{\left(z\right)}$, can be expressed through
a suitable foliation, so that in Fefferman--Graham form (see for instance \cite{Skenderis:2002wp}) the metric reads
\begin{align}
\frac{1}{\ell^2}\,\extd s^{2} & = \frac{\extd r^{2}}{r^{2}}+\frac{\mathcal{L}_{+}}{4}\left(\extd\tilde{x}^{+}\right)^{2}+\frac{\mathcal{L}_{-}}{4}\left(\extd\tilde{x}^{-}\right)^{2}-\left(\frac{r^{2}}{\ell^{2}}+\frac{\ell^{2}\mathcal{L}_{+}\mathcal{L}_{-}}{16r^{2}}\right)\extd\tilde{x}^{+}\extd\tilde{x}^{-}\;,\label{BTZKdV-1}
\end{align}
with the boundary lightcone coordinates
\begin{equation}
\extd\tilde{x}^{\pm}=\mu_{\pm}\extd t\pm \extd\varphi\;,
\end{equation}
where the Lagrange multipliers correspond to $\mu_{\pm}=\delta H_{\pm}^{\left(z\right)}/\delta{\cal L}_{\pm}$,
so that for the BTZ black hole they become
\eq{
\mu_{\pm}=\mathcal{L}_{\pm}^{\frac{z-1}{2}}\;.
}{eq:angelinajolie}
In a holographic context the Lagrange multipliers $\mu_\pm$ are also known as ``sources'' or ``chemical potentials'', and a relation between them and the charges (or vevs) like \eqref{eq:angelinajolie} is known as ``multi-trace deformation'' \cite{Witten:2001ua} (see also \cite{Papadimitriou:2007sj} and references therein). In this language, the KdV bc's summarized above generate a specific 1-parameter family of multi-trace deformations of the usual AdS$_3$/CFT$_2$ setup, 
labeled by the dynamical exponent $z$, with $z=1$ corresponding to the undeformed theory\footnote{A different kind of multi-trace deformations in the context of generalized Gibbs ensembles has been considered in \cite{deBoer:2016bov}.}.

The role of the functionals $H_{\pm}^{\left(z\right)}$ turns out to be extremely relevant, since they do not only define the bc's, but also describe the energies of the left and right movers, which are generically non-chiral. Indeed, left and right energies for configurations that fulfill KdV-type of bc's are given by
\eq{
E_{\pm}^{\left(z\right)}=\frac{\ell}{32\pi G}H_{\pm}^{\left(z\right)},
}{eq:nolabel}
which for BTZ black holes reduce to
\begin{equation}
E_{\pm}^{\left(z\right)}=\frac{\ell}{8G}\frac{1}{z+1}\mathcal{L}_{\pm}^{\frac{z+1}{2}}\,.\label{eq:EmnBTZ-z}
\end{equation}
Note that for the Brown--Henneaux and soft hairy bc's, labeled by $z=1$ and $z=0$, respectively, left and right energies in \eqref{eq:EmnBTZ-z} reduce to the corresponding ones $E_{\pm}^{\left(1\right)}$ and $E_{\pm}^{\left(0\right)}$ in Eqs.~\eqref{eq:BH-Energy} and \eqref{eq:SH-Energy}.

The log corrections to the entropy of BTZ black holes with KdV-type bc's can then also be obtained from the ones with Brown--Henneaux bc's in \eqref{microentropy-1} through the relationship between the microscopic density of states in both ensembles, which now reads
\begin{equation}
\rho\left[E_{\pm}^{\left(z\right)}\right]=\frac{\partial E_{\pm}^{\left(1\right)}}{\partial E_{\pm}^{\left(z\right)}} \, \rho\left[E_{\pm}^{\left(1\right)}\right]\,.\label{eq:jacoibin-2}
\end{equation}
Analogously, by virtue of Eq.~\eqref{microentropy-1}, the logarithm of the microscopic density of states \eqref{eq:jacoibin-2} is now given by 
\begin{equation}
S_{\pm}^{\left(z\right)}=S_{\pm}-\tfrac{3}{2}\ln S_{\pm} +\ln\frac{\partial E_{\pm}^{\left(1\right)}}{\partial E_{\pm}^{\left(z\right)}} \,.
\end{equation}
Thus, making use of the expressions for left and right energies in both ensembles, given in Eqs.~\eqref{eq:BH-Energy} and \eqref{eq:EmnBTZ-z}, the leading term of the black hole entropy and its log corrections read 
\begin{equation}
S_{\text{mic}}^{\left(z\right)}=S_{+}+S_{-}-\left(z+\tfrac{1}{2}\right)\ln\big(S_{+}S_{-}\big)\;.\label{eq:microentropy-z}
\end{equation}
Hence, following Eq.~\eqref{eq:log2}, the log corrections to black hole entropy with bc's of KdV-type are determined by 
\eq{
q_{\textrm{\tiny KdV}} = 2z+1
}{eq:kdv}
Note that the log corrections for Brown--Henneaux and soft hairy bc's, given by $S_{\text{mic}}^{\left(1\right)}$ in \eqref{microentropy-1} and by $S_{\text{mic}}^{\left(0\right)}$ in \eqref{eq:microentropy-0}, are then recovered from $S_{\text{mic}}^{\left(z\right)}$ in \eqref{eq:microentropy-z} for $z=1$ and $z=0$, respectively.

The log corrections for the entropy of BTZ black holes with KdV-type bc's can also be recovered in an alternative form, provided that $z>0$. Indeed, in this case the desired result can be obtained directly from \eqref{eq:Landau} by virtue of the relationship between $E_{\pm}^{\left(z\right)}$ and $T_{\pm}$ given by \cite{Perez:2016vqo} 
\begin{equation}
E_{\pm}^{\left(z\right)}=\frac{\ell}{8G}\,\frac{\left(2\pi\right)^{1+\frac{1}{z}}}{z+1}\,T_{\pm}^{1+\frac{1}{z}}
\end{equation}
so that the entropy in the microcanonical ensemble in Eq.~\eqref{eq:Landau} precisely reduces to \eqref{eq:microentropy-z}.

The correct result for the log corrections in the case of soft hairy bc's is recovered in the limit $z\to 0^+$, which naturally suggests that a small positive dynamical exponent $z$ can be regarded as a regulator providing a (positive) finite specific heat. In the next section we exploit this suggestion in order to carry out a microscopic counting.

\section{Microscopic counting from anisotropic S-duality \label{sec:Microscopic-log-corrections}}

In microscopic derivations of log corrections a technical tool of relevance is a continuous analog of Kramers--Wannier high/low temperature duality \cite{Kramers:1941kn}, usually referred to as S-duality, to evaluate the partition function and its correction from Gaussian fluctuations. 

The soft hairy bc's are special in this sense because,
as pointed out in the introduction, the global charges, and in particular
the energy, do not depend on the temperature, which spoils any chance
of establishing a high/low temperature duality. This goes hand in hand
with the anisotropic scaling properties they possess, since there
is no natural notion of modular invariance, because Euclidean time
does not scale for $z=0$. Nevertheless, the moral learned from the
previous section strongly points towards deforming the theory by introducing
a generic dynamical exponent $z$ which can be regarded as a regulator,
so that the leading term of the entropy and the log corrections
could be obtained in the limit $z\to 0^+$. Indeed, this strategy
has already been successfully implemented for the leading term of
the entropy in \cite{Afshar:2016kjj} in order to recover the correct
result for the Bekenstein-Hawking entropy in the case of soft hairy
bc's.

\subsection{Asymptotic growth of states for dual theories with Lifshitz type scaling}

Here we briefly summarize and exploit an anisotropic version of S-duality that holds for dual theories possessing anisotropic scaling of Lifshitz type with dynamical exponent $z$ along the lines of \cite{Gonzalez:2011nz,Perez:2016vqo}.\footnote{%
Alternative extensions of S-duality have also been discussed for different
scaling laws in \cite{Barnich:2012xq,Bagchi:2012xr,Detournay:2012pc,Shaghoulian:2015dwa,Bravo-Gaete:2015wua,Castro:2015uaa,Detournay:2016gao}.} 
Thus, we consider two-dimensional field theories described by two decoupled left and right movers with the same anisotropic scaling of time ($t$) and angular ($\varphi$) coordinates 
\begin{equation}
t\to\lambda^{z}t\text{\quad\quad}\varphi\to\lambda\varphi\;,\label{eq:log3}
\end{equation}
with arbitrary (not necessarily integer) dynamical exponent $z>0$.

Its Euclidean continuation is then defined on a torus characterized
by a modular parameter $\tau$. As shown in \cite{Gonzalez:2011nz,Perez:2016vqo},
the generalization of S-duality in the anisotropic case is given by
the following high/low temperature relationship 
\begin{equation}
\tau\to i^{1+\frac{1}{z}}\tau^{-\frac{1}{z}}\;,\label{eq:Lifmodular}
\end{equation}
so that the partition function is assumed to be invariant under 
\begin{equation}
Z\left[\tau;z\right]=Z\left[i^{1+\frac{1}{z}}\tau^{-\frac{1}{z}};z^{-1}\right]\:.\label{eq:lalapetz}
\end{equation}
Note that for $z=1$ the duality \eqref{eq:lalapetz} reduces to standard S-duality as part of modular invariance in a CFT$_2$ \cite{Cardy:1986ie,DiFrancesco}. 

Under the assumption of the existence of a gap in the spectrum, at
low temperatures the ground state dominates the partition function.
If the ground state is non-degenerate with negative left and right
energies $-\Delta_{0}^{\pm}\left[z\right]$ that generically depend
on $z$, then at low temperatures the partition function approximates
as 
\begin{equation}
Z[\tau;z]\approx e^{-2\pi i\left(\tau\Delta_{0}\left[z\right]-\bar{\tau}\bar{\Delta}_{0}\left[z\right]\right)}\;.
\end{equation}
Using the anisotropic S-duality relation \eqref{eq:Lifmodular}, at
high temperatures the partition function then behaves as 
\begin{equation}
Z[\tau;z]\approx e^{2\pi\left(\left(-i\tau\right)^{-\frac{1}{z}}\Delta_{0}\left[z^{-1}\right]+\left(i\bar{\tau}\right)^{-\frac{1}{z}}\bar{\Delta}_{0}\left[z^{-1}\right]\right)}\;.\label{eq:Z-Modu-2}
\end{equation}

On the other hand, the inverse Laplace transform allows to obtain the asymptotic growth of the number of states at fixed energy $\Delta\gg\Delta_{0}$ according to 
\begin{equation}
\rho\left[\Delta,\bar{\Delta}\right]=\int\extd\tau\extd\bar{\tau}\,Z[\tau;z]e^{-2\pi i\left(\tau\Delta-\bar{\tau}\bar{\Delta}\right)}\approx\int\extd\tau \extd\bar{\tau}\,e^{f\left(\tau,\Delta\right)}e^{\bar{f}\left(\bar{\tau},\bar{\Delta}\right)}
\end{equation}
with 
\begin{equation}
f\left(\tau,\Delta\right):=2\pi\left(\left(-i\right)^{-\frac{1}{z}}\Delta_{0}\left[z^{-1}\right]\tau^{-\frac{1}{z}}-i\Delta\tau\right)\:.
\end{equation}
This expression can be readily evaluated in the saddle point approximation.
The equilibrium value of $\tau$ that extremizes $f\left(\tau,\Delta\right)$
is given by 
\begin{equation}
\tau_{*}=i\big(z\Delta\big)^{-\frac{z}{z+1}}\big(\Delta_{0}\left[z^{-1}\right]\big)^{\frac{z}{z+1}}\;.
\end{equation}
Expanding $f\left(\tau,\Delta\right)$ around this equilibrium up to terms of cubic order yields 
\begin{align}
\rho\left[\Delta,\bar{\Delta}\right] & \approx e^{f\left(\tau_{*},\Delta\right)+\bar{f}\left(\bar{\tau}_{*},\bar{\Delta}\right)}\int\extd\tau \extd\bar{\tau}\,e^{\frac{1}{2}\left(\tau-\tau_{*}\right)^{2}\partial_{\tau}^{2}f\left(\tau_{*},\Delta\right)}e^{\frac{1}{2}\left(\bar{\tau}-\bar{\tau}_{*}\right)^{2}\partial_{\bar{\tau}}^{2}\bar{f}\left(\bar{\tau}_{*},\bar{\Delta}\right)}\;.
\label{eq:gauss}
\end{align}
Therefore, performing the Gaussian integrals \eqref{eq:gauss}, the asymptotic growth of the number of states is given by
\begin{equation}
S_{\text{mic}}^{\left(z\right)}=\ln\rho\left[\Delta,\bar{\Delta}\right]=S^{E}-\left(z+\tfrac{1}{2}\right)\,\ln S^{E}+c.c.+\dots
\end{equation}
with
\begin{equation}
S^{E}=2\pi\left(1+z\right)\Delta^{1/\left(1+z\right)}\exp\left[\tfrac{z}{1+z}\ln\left(\Delta_{0}\left[z^{-1}\right]/z\right)\right]\;.
\end{equation}

In terms of Lorentzian left and right energies $\Delta_\pm$ the result is given by 
\begin{equation}
S_{\text{mic}}^{\left(z\right)}=S_{+}+S_{-}-\left(z+\tfrac{1}{2}\right)\ln\big(S_{+}S_{-}\big)+\dots\label{microentropy}
\end{equation}
with \cite{Perez:2016vqo} 
\begin{equation}
S_{\pm}=2\pi\left(1+z\right)\Delta_{\pm}^{1/\left(1+z\right)}\exp\left[\tfrac{z}{1+z}\ln\left|\Delta_{0}^{\pm}\left[z^{-1}\right]/z\right|\right]\;.\label{eq:whatever}
\end{equation}
Note that the leading term of the entropy for each of the movers acquires the expected dependence on the energy for a field theory with Lifshitz scaling in two dimensions, see e.g.~\cite{Taylor:2008tg, Bertoldi:2009vn, Bertoldi:2009dt, Hartnoll:2009sz, DHoker:2010zpp, Hartnoll:2011fn, Hartnoll:2015faa, Taylor:2015glc}. The presence of the ground state energy with inverse dynamical exponent, $\Delta_{0}^{\pm}\left[z^{-1}\right]$, requires acquired taste. However, there are two cases where this issue is absent: either $z=1$ (Brown--Henneaux bc's) in which case $\Delta_{0}^{\pm}\left[z^{-1}\right]=\Delta_{0}^{\pm}\left[z\right]$, or $z=0$ (soft hairy bc's) in which case the whole exponent in \eqref{eq:whatever} simplifies to unity due to the prefactor linear in $z$ so that the only information required from $\big|\Delta_{0}^{\pm}\left[z^{-1}\right]/z\big|$ is that it has no essential singularity as $z\to 0^+$.

\subsection{Black hole entropy and log corrections for KdV-type boundary conditions}

As explained in \cite{Perez:2016vqo}, once expressed in terms the
extensive variables, given by left and right energies $E_{\pm}^{\left(z\right)}$,
the Bekenstein-Hawking entropy of BTZ black holes that fulfill bc's of KdV-type reads
\begin{equation}
S_{\text{BH}}=\frac{A}{4G}=\frac{\pi\ell}{4G}\left(\tfrac{8G}{\ell}\left(z+1\right)\right)^{\frac{1}{z+1}}\left[\left(E_{+}^{\left(z\right)}\right)^{\frac{1}{z+1}}+\left(E_{-}^{\left(z\right)}\right)^{\frac{1}{z+1}}\right]\;.\label{eq:EntropyBTZ-KdV-BCs}
\end{equation}
The black hole entropy in \eqref{eq:EntropyBTZ-KdV-BCs}
is successfully recovered from the leading term of \eqref{microentropy}
provided that left and right energies of the black hole, $E_{\pm}^{\left(z\right)}$,
are identified with the corresponding ones of the dual theory with
anisotropic scaling, $\Delta_{\pm}$, while the ground state energies
of the dual theory, $\Delta_{0}^{\pm}\left[z\right]$, correspond
to the ones of global AdS$_3$, given by \eqref{eq:EmnBTZ-z}
evaluated for $\mathcal{L}_{\pm}=-1$ . Indeed, it is simple to verify
that if one inserts the KdV left and right energies
\begin{equation}
\Delta_{\pm}=E_{\pm}^{\left(z\right)}=\frac{\ell}{8G}\frac{1}{z+1}\mathcal{L}_{\pm}^{\frac{z+1}{2}}\,,\label{eq:Deltamn-z}
\end{equation}
and the corresponding ground state energies
\begin{equation}
\Delta_{0}^{\pm}\left[z\right]=\frac{\ell}{8G}\frac{1}{z+1}\left(-1\right)^{\frac{z+1}{2}}\,,\label{eq:Delta0mn-z}
\end{equation}
into the general result for the entropy formula \eqref{eq:whatever}, one precisely recovers the leading term of the black hole entropy, since 
\eq{
S_{+}+S_{-}=\frac{\ell\pi}{4G}\left(\sqrt{{\cal L}_{+}}+\sqrt{{\cal L}_{-}}\right)=\frac{A}{4G}\;.
}{eq:yetanother}

Consequently, it is reassuring to verify that identifying left and
right energies of the black hole with the ones of the dual theory
as in \eqref{eq:Deltamn-z}, \eqref{eq:Delta0mn-z}, not only allows
to successfully reproduce the leading term of the entropy, but actually
also yields the correct log corrections from anisotropic
S-duality, given by
\begin{equation}
q=2z+1\;.
\label{eq:q}
\end{equation}
The result \eqref{eq:q} agrees with the log corrections \eqref{eq:kdv}, which were obtained from an entirely different approach. 

Note in particular that for $z=1$ the isotropic result for the black hole entropy is recovered, both for the leading term \cite{Strominger:1997eq} as well as for the log corrections \cite{Carlip:2000nv, Sen:2012dw}, where the role of the central charge is played by the energy of global AdS$_3$.

\subsection{Black hole entropy and log corrections for soft hairy boundary conditions}\label{subsec:Black-hole-entropy}

In the case of soft hairy bc's \cite{Afshar:2016wfy} it has been shown that the black hole entropy depends only on the zero modes of the affine $\hat{u}\left(1\right)$ charges $J_{0}^{\pm}=E_{\pm}^{\left(0\right)}$, i.e.,
\begin{equation}
S_{\text{BH}}=\frac{A}{4G}=2\pi\left(J_{0}^{+}+J_{0}^{-}\right)\;.\label{eq:EntropySoft-BH}
\end{equation}

As pointed out at the beginning of this section, our strategy to recover the black hole entropy and its log corrections in this case ($z=0$) consists in regarding the microscopic counting performed above where the dynamical exponent $z$ is considered as a small positive regulator. Therefore, left and right energies of the black hole are identified with the ones of the dual theory with anisotropic scaling, i.e.,
\begin{equation}
\Delta_{\pm}=E_{\pm}^{\left(0\right)}=J_{0}^{\pm}\,.\label{eq:Deltamn-z0-J0}
\end{equation}
In this step, it is worth pointing out that global AdS$_{3}$
is not included within the soft hairy bc's (for real $J_0^\pm$). Nevertheless,
since the soft hairy bc's correspond to the KdV-type
ones for $z=0$, in the regularization process, before taking the
limit $z\to 0^+$, it is natural to assume that the ground state
energies of the dual theory correspond to the ones of AdS$_{3}$ with
bc's of KdV-type with $z\neq$0, given by \eqref{eq:Delta0mn-z},
so that
\begin{equation}
\Delta_{0}^{\pm}\left[z^{-1}\right]=\frac{\ell}{8G}\frac{z}{z+1}\left(-1\right)^{\frac{z+1}{2z}}\,.\label{eq:Delta0mn-zm1}
\end{equation}
This means that the quantity $|\Delta_{0}^{\pm}\left[z^{-1}\right]/z|=\tfrac{\ell}{8G}\,\tfrac{1}{z+1}$ is regular as $z\to 0^+$. Plugging the expressions \eqref{eq:Deltamn-z0-J0} and \eqref{eq:Delta0mn-zm1} into the microcanonical KdV entropy \eqref{microentropy}, \eqref{eq:whatever} establishes
\begin{equation}
S_{\text{mic}}^{\left(0\right)}:=\lim_{z\to 0^+}S_{\text{mic}}^{\left(z\right)}=S_{+}+S_{-}-\tfrac{1}{2}\ln\big(S_{+}S_{-}\big)
\end{equation}
with 
\begin{equation}
S_{\pm}=2\pi J_{0}^{\pm}\;.
\end{equation}
The leading term as well as the log corrections ($q_{\textrm{\tiny sh}}=1$) of the black hole entropy with soft hairy bc's precisely agree with results found in section \ref{subsec:Soft-hairy-and}, given by Eq.~\eqref{eq:microentropy-0}, which were found from a radically different approach. This suggests that the introduction of a small positive dynamical exponent $z$ as regulator is appropriate and the limit $z\to 0^+$ meaningful.

As a closing remark of this section we would like mentioning that our result for the log corrections in the case of $z=0$ can also be recovered in an alternative way as in \cite{Afshar:2017okz}, which follows the lines of the ``horizon fluff'' proposal \cite{Afshar:2016uax, Sheikh-Jabbari:2016npa}, thereby providing semi-classical support for this proposal.

\section{Log corrections for the entropy of soft hairy black holes}\label{sec:Logarithmic-corrections-for-soft-hairy}

In sections \ref{subsec:Soft-hairy-and} and \ref{subsec:Black-hole-entropy} it has been shown from different approaches that the asymptotic growth of the number of states for BTZ black holes that fulfill soft hairy bc's is given by
\begin{equation}
\rho\left[J_{0}^{+},\,J_{0}^{-}\right]=\rho\left[J_{0}^{+}\right]\rho\left[J_{0}^{-}\right]\label{eq:rhoJmJn}
\end{equation}
with
\begin{equation}
\rho\left[J_{0}^{\pm}\right]= \frac{e^{2\pi J_{0}^{\pm}}}{\sqrt{2\pi J_{0}^{\pm}}}\,.
\label{eq:rhomnJ0mn}
\end{equation}
Nonetheless, the soft hairy bc's in \cite{Afshar:2016wfy}
also include stationary non-spherically symmetric black holes that
carry all of the possible affine charges. It is then natural to wonder
about the asymptotic growth of the number of states for this class
of soft hairy black holes. Remarkably, if one assumes that the chemical
potentials associated to soft hair charges $J_{n}^{\pm}$, with $n\neq0$,
are switched off, it is possible to show that not only the leading
term, but also the log correction for the entropy of this
set of black flowers coincide with the ones described by \eqref{eq:rhoJmJn}
and \eqref{eq:rhomnJ0mn}. This can be seen as follows.

The global charges that soft hairy black holes are endowed with correspond
to the generators of two independent copies of $\hat{u}\left(1\right)$,
which span the asymptotic symmetries. The simplest configuration only
carries zero mode charges which determine the values of the outer
and inner horizons. Indeed, this configuration is spherically symmetric
and corresponds to the BTZ black hole. It is then worth pointing out
that the generic soft hairy black hole solution can be generated from
the BTZ black hole through the application of a generic ``soft boost'',
which consists of acting on the spherically symmetric solution with
any set of affine asymptotic symmetry generators. The ``softly boosted'' solutions remain regular, which goes in hand
with the fact that the chemical potentials associated with soft hair
charges vanish. Importantly, this process
does not change the value of the zero modes, since they commute with
all the remaining soft hairy global charges, i.e.,
\begin{equation}
\left[J_{0}^{\pm},J_{n}^{\pm}\right]=0\,.
\end{equation}

Therefore, by virtue of the process aforementioned one finds that the entropy of soft hairy black holes, including the log corrections, depends only on the zero modes and coincides with the one found for BTZ black holes.

\section{Outlook to generalizations}\label{sec:outlook}

Remarkably, it was shown in \cite{Grumiller:2016kcp} that the right-hand side of \eqref{eq:EntropySoft-BH},
\eq{
S=2\pi\,\big(J_0^+ + J_0^-\big)
}{eq:shs}
also reproduces the entropy of black holes (as well as the entropy of flat space cosmologies \cite{Ammon:2017vwt}) in the case of higher spin gravity in three spacetime dimensions \cite{Perez:2012cf,Perez:2013xi,deBoer:2013gz,Bunster:2014mua}, which differs from the Bekenstein--Hawking result \eqref{eq:log1}. It could be interesting to calculate leading order expressions and log corrections to the entropy of higher spin black holes and (spin-2 or higher spin) flat space cosmologies along the lines of the present work, which could be compared with the approach in \cite{Gaberdiel:2012yb}.

Other possible generalizations depend on the universality of the soft hairy entropy formula \eqref{eq:shs} (or its higher-dimensional counterpart). For instance, it could be of interest to consider generalizations to supergravity, to black hole solutions that are not maximally symmetric, to massive gravity theories and/or to black holes in four or higher dimensions. The first step towards these generalizations is to recover (or generalize) the soft hairy entropy formula \eqref{eq:shs}. 
Perhaps the simplest generalization is to consider flat space cosmologies in three dimensions, whose log corrections to their Bekenstein--Hawking entropy was calculated in \cite{Bagchi:2013qva}. Since there is a soft hairy version of flat space cosmologies \cite{Afshar:2016kjj} it seems likely that the results of the current paper generalize to this case.

Finally, it could be rewarding to consider extremal or near-extremal limits of our results and of the bc's introduced in \cite{Afshar:2016wfy}. In the near extremal case one of the chiral entropies $S_\pm$ is order of unity, say $S_-\sim{\cal O}(1)$, and the result \eqref{microentropy-1} for Brown--Henneaux bc's is modified to $S_{\textrm{\tiny mic}}^{\left(1\right)}\big|_{\textrm{\tiny ext}}=S_{+}-\tfrac{3}{2}\ln S_{+}+\dots$, so that effectively the value of the numerical coefficient $q_{\textrm{\tiny ext}}$ in front of the log correction is half the non-extremal one, $q_{\textrm{\tiny ext}}=\tfrac32$. 

\acknowledgments

We thank Hamid Afshar, Hern\'an Gonz\'alez, Marc Henneaux, Wout Merbis, Edgar Shaghoulian, Shahin Sheikh-Jabbari, Andrew Strominger and Hossein Yavartanoo for discussions. DG and RT are grateful for the stimulating atmosphere of the focus week \textit{Recent developments in AdS$_{3}$ black-hole physics} within the program ``New Developments in AdS$_{3}$/CFT$_{2}$ Holography'' in April 2017 at GGI in Florence where this work was completed. DG also acknowledges the hospitality at CECs in November 2016, where this project was started.

The work of DG was supported by the Austrian Science Fund (FWF), projects P~27182-N27 and P~28751-N27. The work of AP, DT and RT is partially funded by Fondecyt grants 11130260, 11130262, 1161311, 1171162. The Centro de Estudios Cient\'ificos (CECs) is funded by the Chilean Government through the Centers of Excellence Base Financing Program of Conicyt.


\providecommand{\href}[2]{#2}\begingroup\raggedright\endgroup

\end{document}